\documentclass[aps,showpacs,onecolumn,twoside]{revtex4}
\usepackage{epsfig,amsmath,amssymb}

\expandafter\ifx\csname package@font\endcsname\relax\else
 \expandafter\expandafter
 \expandafter\usepackage
 \expandafter\expandafter
 \expandafter{\csname package@font\endcsname}%
\fi

\def\ket#1{|#1\rangle}
\def\bra#1{\langle#1|}

\begin{document}



\title{\bf{PERES LATTICES IN NUCLEAR STRUCTURE}}

\author{\footnotesize Michal Macek, Pavel Str\'ansk\'y, Pavel Cejnar}

\affiliation{Faculty of Mathematics and Physics, Charles University, V Hole\v{s}ovi\v{c}k\'ach 2,\\
180 00 Prague,
Czech Republic}



\begin{abstract}  
A method by Peres is used to draw spectra of the interacting boson model as lattices in the plane of energy versus an arbitrary 
observable average (or variance).
Ordered (disordered) lattices are signatures of regularity (chaos) in both quantum and classical dynamics. 
The method is also apt to disclose exact or approximate dynamical symmetry.
\end{abstract}


\maketitle


\vline
  
We use a method by Asher Peres~\cite{ref:Per84} to study order and chaos within the spectra of collective nuclear models. The method allows to distinguish order/chaos in a quantum system by visual inspection, which is in a way similar to the well-known Poincar\'e section method used in classical mechanics. In contrast to the conventional methods of quantum chaology 
 based mostly on statistical properties of energy eigenvalues found in a certain interval, the lattice method enables one to assign regular or chaotic type of dynamics to \emph{individual} states. 

Within every quantum-mechanical model, we can construct trivial integrals of motion---we call them Peres invariants $\hat{P}(\hat{O})$---by taking the time average of an arbitrary ``well-behaved'' operator $\hat{O}$:
\begin{equation}\label{eq:Peres}
\hat{P}(\hat{O}) = \lim_{T\rightarrow\infty}\frac{1}{T}\int^T_0\hat{O}(t)dt\,,
\end{equation}
where $\hat{O}(t)$ is the Heisenberg image of $\hat{O}$. By plotting the expectation values $P_i(\hat{O})=\bra{\psi_i}\hat{P}(\hat{O}) \ket{\psi_i}$ versus the energy $E_i = \bra{\psi_i} \hat{H} \ket{\psi_i}$ for an arbitrary set of states $\ket{\psi_i}$, we obtain what we call Peres lattices. If $\ket{\psi_i}$ are eigenstates of the Hamiltonian $\hat{H}$, the expression $\bra{\psi_i}\hat{P}(\hat{O}) \ket{\psi_i}$ can be simply replaced by $\bra{\psi_i} \hat{O} \ket{\psi_i}$.

A regular lattice of points can be expected, if the system is integrable~\cite{ref:Per84}. The supporting arguments are based on semiclassical Einstein-Brillouin-Keller (EBK) quantisation: For a classically integrable system in $d$ dimensions, the phase space trajectories are bound to $d$-dimensional tori. A complete set of motion integrals can be formed by the action variables $J_k = \oint_{C_k}p_i dx^i$, where $i,k=1,..,d$ and $C_k$ are topologically non-equivalent curves on the torus surfaces. From the EBK quantisation we obtain $J_k = \hbar n_k + a_k$, where $a_k$ are constants and $n_k = 0,1,2,..$integers. Any other integral of motion---including energy $E$ and the Peres invariant $\hat{P}(\hat{O})$---can be expressed as a smooth function of $J_k$. Hence follows that the lattice of $P(\hat{O})$ versus $E$ is a smooth transform of the regular lattice $J_k$ versus $E$ and should in most cases be visually regular.

Adding a perturbation to an integrable system brings about formation of irregular patterns in the Peres lattices. For weak perturbations, the regular lattice usually does not break down ``uniformly''. On the contrary, chaos develops in localised segments of the lattice, while the rest may remain untouched, as seen from the studies within the geometric collective model and interacting boson model of nuclei~\cite{ref:StrCGS,ref:StrMacXX}. The method allows to select the states most affected by the perturbation and is thus appealing for studies of the onset of quantum chaos. 

To demonstrate the essence of the method, we consider the interacting boson model (IBM) Hamiltonian~\cite{ref:IachBook} in the simple and effective form 
	\begin{equation}
		H=\eta\,\hat{n}_{d}-(1-\eta)\,N^{-1}\,\hat{Q}^{\chi}\cdot\hat{Q}^{\chi},	
		\label{eq:H}
	\end{equation}
	with $n_{d}= d^{\dag}\cdot\tilde{d}$ the $d$-boson number operator,
	$\hat{Q}_{\mu}^{\chi} = d^{\dag}_{\mu}s + s^{\dag}\tilde{d}_{\mu} + \chi[d^{\dag}\tilde{d}]^{(2)}_{\mu}$ the quadrupole operator, $N$ the total number of bosons being conserved and $(\eta,\chi)$ two external parameters defining the Casten triangle.
	Its vertices $(\eta,\chi)=(1,0)$, $(0,0)$, and $(0,-\sqrt{7}/2)$ correspond to the U(5), O(6), and SU(3) dynamical 
	symmetries, respectively.

	\begin{figure}[ht]
	\includegraphics[width=\textwidth]{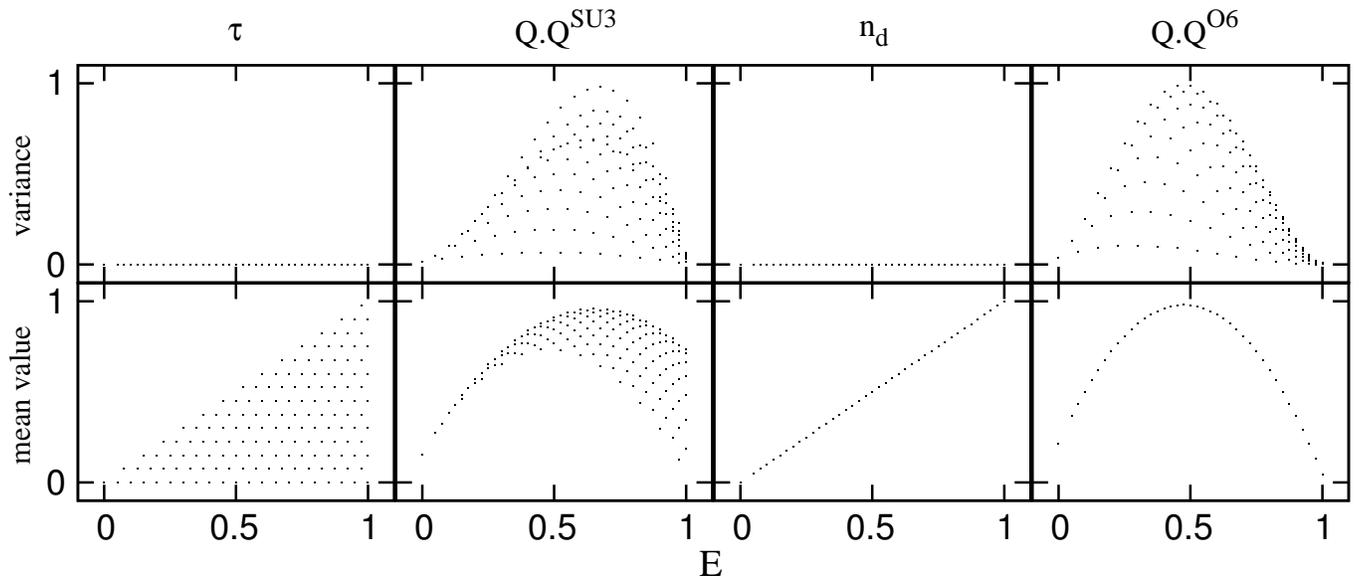}
	\caption{
		Peres lattices for selected invariants calculated in the integrable U(5) limit of IBM (bottom row) for $L=0$ states and $N=40$ bosons. The corresponding variance lattices (top row) show non-zero values for $\langle \hat{Q}.\hat{Q}^{\mathrm{SU(3)}}\rangle$ and $\langle \hat{Q}.\hat{Q}^{\mathrm{O(6)}}\rangle$ and zero for the good quantum numbers $\tau$, $n_d$. All quantities are scaled by their maximum value to fit in the interval $[0,1]$.   	
	}
	\label{fig:U5}
	\end{figure}
		
The striking property of the Peres method---the arbitrariness of choice of $\hat{O}$ in Eq.~(\ref{eq:Peres})---is demonstrated here in the completely integrable U(5) limit of~(\ref{eq:H}). In the bottom row of Fig.~\ref{fig:U5}, we plot the Peres lattices related to the Casimir operators of groups O(5), SU(3), U(5) and O(6). Notably, \emph{all} quantities display completely regular lattices despite the fact that only $\tau$ (O(5) label) and $n_d$ (U(5) label) are indeed exact quantum numbers. Non-existence of quantum numbers related to $\hat{Q}.\hat{Q}^{\mathrm{SU(3)}}$ and $\hat{Q}.\hat{Q}^{\mathrm{O(6)}}$ is evident from the variance lattices shown above the corresponding Peres invariants---all eigenstates display non-zero value of the variance $\mathrm{var}(\hat{O})=\langle\hat{O}^2\rangle - \langle\hat{O}\rangle^2$. 

\begin{figure}[ht]
  \includegraphics[width=\textwidth]{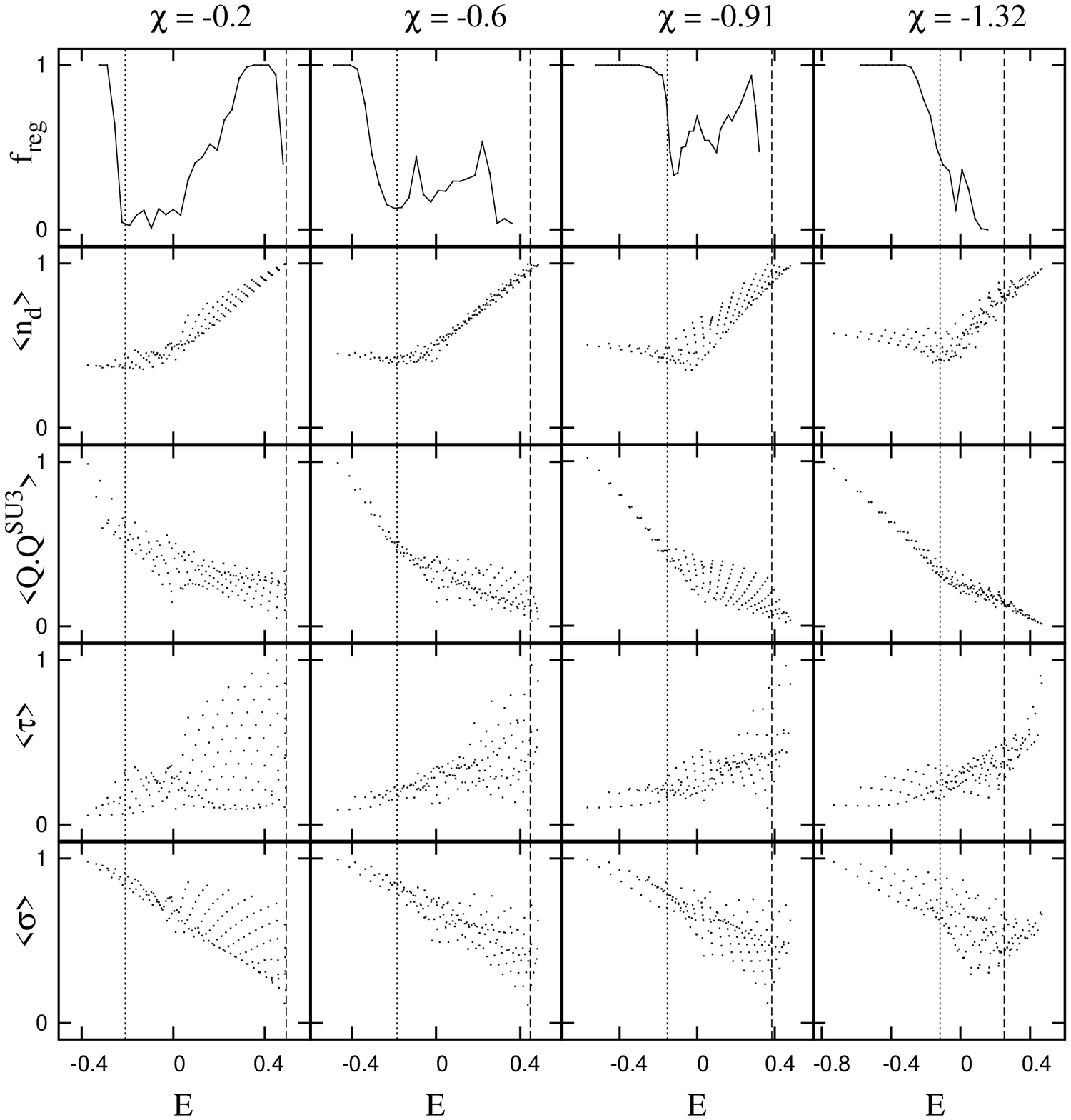}
	\caption{
		Peres lattices of $L=0$ eigenstates of the IBM Hamiltonian (\ref{eq:H}) for U(5) and SU(3) Casimir invariants and O(6) and O(5) labels $\sigma$, $\tau$ and the corresponding classical regular fraction $f_\mathrm{reg}$ calculated inside the Casten triangle for $N=40$ bosons. The parameter $\eta=0.5$ and $\chi$ is indicated on the top of individual columns. The dotted and dashed lines mark respectively $E_\mathrm{sad}$ and $E_{\lim}$ of the classical IBM potential. Each invariant is scaled to fit in the interval $[0,1]$.
	}
	\label{fig:Inside}
	\end{figure}
			
Having tested the Peres' proposal in one of the integrable cases, we can explore the interior of the Casten triangle, which is known to display mixed regular/chaotic dynamics~\cite{ref:Alh,ref:Arc}. Figure~\ref{fig:Inside} illustrates the evolution of various Peres invariants---$n_d$, $\hat{Q}.\hat{Q}^{\mathrm{SU(3)}}$, $\tau$ and $\sigma$ (O(6) label)---for selected values of $\chi$ changing across the Casten triangle at $\eta=0.5$. Panels devoted to a particular invariant are arranged in rows, while the columns correspond to a given value of $\chi$. In the top row, we show the classical regularity expressed by the regular fraction $f_{\mathrm{reg}}$ of the phase space (see Ref.~\cite{ref:Arc} for details).
	
Peaks of $f_{\mathrm{reg}}$ correspond to distinct regular areas in the lattices of all the Peres invariants considered. In the energy intervals of mixed dynamics, we may point out states belonging to the regular lattices. A thorough analysis shows that for every Peres invariant the ``regular'' states form a completely regular lattice if plotted separately without the remaining ``chaotic'' states. The regular states correspond to resilient classical tori~\cite{ref:StrMacXX}. 

As the energy $E$ increases, we observe changes of the form and pattern in the lattices. Major changes in the overall form are noticeably triggered by stationary points of the classical IBM potential (see Ref.~\cite{ref:Arc})---the saddle point energy $E_\mathrm{sad} $(dotted vertical lines in Fig.~\ref{fig:Inside}), the local maximum at $E=0$ and the limiting energy for the system to be bound $E_{\lim}$(dashed lines). Probably the most significant overall feature is the ``linear'' dependence of $\langle \hat{Q}.\hat{Q}^{\mathrm{SU(3)}}\rangle$ on $E$ below $E_{\mathrm{sad}}$ roughly in the range of $\chi < -0.8$, which is connected with the quasi SU(3) dynamical symmetry~\cite{ref:StrMacXX}.  
	

The lattice method is applicable for any quantum model, possibly beyond nuclear structure physics. It is in particular well suited to disclose approximate symmetries of the model, like quasi dynamical symmetry~\cite{ref:RoweQDS} and partial dynamical symmetry~\cite{ref:LeviPDS}, especially using the variance lattices.
More results concerning the dynamics of both the interacting boson model and the geometric collective model are accessible in an interactive way at our web site~\cite{ref:www} and will be described in deeper detail in our prepared publications~\cite{ref:StrMacXX}.

This work was supported by the Czech Science Foundation (202/06/0363) and by the Czech Ministry of Education (MSM 0021620859 and LA 314).

\end{document}